\documentclass[pre,aps,twocolumn,amsmath,amssymb,showpacs]{revtex4}
\usepackage{graphicx}
\usepackage{epsfig}
\include{graphics}
\begin{document}

\title{Monte Carlo simulations of pulse propagation in massive 
multichannel optical fiber communication systems}

\author{Yeojin Chung}
\affiliation{Department of Mathematics, Southern Methodist University, 
Dallas, Texas 75275, USA}

\author{Avner Peleg}
\affiliation{Department of Mathematics, State University of New York
at Buffalo, Buffalo, New York 14260, USA}


\begin{abstract}
We study the combined effect of delayed Raman response and 
bit pattern randomness on pulse propagation in massive 
multichannel optical fiber communication systems. 
The propagation is described by a perturbed stochastic 
nonlinear Schr\"odinger equation, which takes into account 
changes in pulse amplitude and frequency as well as emission of 
continuous radiation. We perform extensive numerical simulations 
with the model, and analyze the dynamics of the frequency moments, 
the bit-error-rate, and the mutual distribution of amplitude and 
position. The results of our numerical simulations are in good 
agreement with theoretical predictions based on the adiabatic 
perturbation approach.               
\end{abstract}

\maketitle

\section{Introduction}
\label{Introduction}
The interplay between noisy phenomena and nonlinear processes 
is a rich field of research that is of great interest in a 
variety of disciplines including solid state 
physics \cite{Malomed89}, turbulence \cite{Frisch95}, 
and optics \cite{Agrawal2001}. One of the most important problems 
in this field concerns propagation of coherent 
patterns, such as solitons and solitary waves, in the presence of 
noise and/or disorder. An excellent example for systems where 
noise and nonlinear effects play an important role in the dynamics 
of coherent patterns is provided by fiber optics communication systems, 
which employ optical pulses to represent bits 
of information \cite{Agrawal2001}. It is by now 
well established that the parameters characterizing 
the pulses in fiber optics communication systems can exhibit 
non-Gaussian statistics 
\cite{Menyuk95,Georges96,Falkovich2001,Biondini2002,Falkovich2004}. 
Yet, since optical fiber systems are only 
weakly nonlinear, it was commonly believed that the statistics of 
optical pulses is very different from the statistics encountered 
in strongly nonlinear systems, such as turbulence and chaotic flow, 
where intermittent dynamics exists. However, a recent study of 
pulse propagation in optical fiber systems with multiple 
frequency channels in the presence of delayed Raman response 
obtained results that stand in sharp contrast to this common 
belief \cite{P2007}. This study focused on the interplay between 
Raman induced energy exchange in pulse collisions and randomness 
of pulse sequences in different frequency channels. 
Taking into account these two effects it was shown that the pulse 
parameters exhibit intermittent dynamic behavior in the sense that 
their normalized moments grow exponentially with propagation distance. 
Furthermore, it was shown that this intermittent dynamic behavior 
has important practical consequences, by leading to relatively 
large values of the bit-error-rate (BER), which is the probability 
for an error at the output of the fiber line.

The results of the study in Ref. \cite{P2007} were based on an adiabatic 
perturbation procedure that neglects radiation emission effects.
However, these effects can be especially important for the fiber optics 
system under consideration. Indeed, as we shall see below, the 
interplay between collision-induced energy exchange and randomness 
of pulse sequences can be described as an effective disorder in the 
linear gain/loss coefficient, and the presence of gain can lead to 
instability with respect to emission of continuous radiation. Therefore,  
it is essential to obtain an improved description of pulse dynamics in 
the system that includes emission of continuous waves. In this paper we 
take this important task and derive a perturbed stochastic nonlinear 
Schr\"odinger (NLS) equation, which takes into account both changes 
in pulse parameters and radiation emission effects. We employ 
this model to analyze the dynamics of soliton parameters 
in comparison with the results of the simpler description 
of Ref. \cite{P2007} and to draw conclusions on the 
possibility to observe intermittent dynamics in multichannel 
optical fiber communication systems. The rest of the introduction 
is devoted to a summary of previous research on the effects of 
delayed Raman response on soliton propagation in optical fibers.

The main effect of Raman scattering on
a single soliton propagating in the fiber is the self frequency shift. 
This effect, which is caused by energy transfer from higher 
frequency components of the pulse to its lower frequency components,   
was first observed experimentally by Mitschke and 
Mollenauer \cite{Mitschke86} and explained 
theoretically by Gordon \cite{Gordon86}. Following this discovery, 
the impact of delayed Raman response on soliton propagation 
in optical fibers has drawn a lot of attention
\cite{Kodama87,Chi89,Malomed91,Stolen92,Agrawal96,Kumar98,Kaup99,
Chung2002,Skryabin2003}. Most significantly, the influence 
on two-soliton collisions was studied by numerical 
simulations \cite{Chi89,Agrawal96} as well as by 
theoretical analysis \cite{Malomed91,Kumar98,Kaup99}. These studies 
revealed that the main effect of a single two-soliton collision in the 
presence of delayed Raman response is an energy exchange between the 
colliding pulses, which leads to a change of their amplitudes 
(Raman induced cross talk) \cite{Chi89,Malomed91,Agrawal96,Kumar98,Kaup99}. 
The frequencies of the two solitons was also found to 
change as a result of the collision (Raman induced cross frequency
shift) \cite{Chi89,Agrawal96,Kumar98,Kaup99}. Similar effects 
were recently studied in collisions of ultra-short 
soliton pulses in photonic crystal fibers \cite{Skryabin2006,Bang2006}.

Raman induced energy exchange in pulse collisions can be 
beneficially employed in a variety of applications, including 
amplification in fiber lines \cite{Islam2004,Agrawal2005} and 
in tunable laser sources \cite{Stolen72,Agrawal2001}. However, it 
can also have negative effects that impose severe limitations on 
the performance of multichannel communication systems. Indeed, it 
is known that the Raman-induced energy exchange in a single 
interchannel collision is independent of the frequency 
difference between the channels. Consequently, 
the magnitude of the induced energy shifts for a given pulse  
grows with the square of the number of channels
\cite{Tkach97,Kaup99}. Thus, in a 100-channel system, 
for example, these effects can be larger by a factor 
of $2.5\times 10^{3}$ compared with a two-channel system  
operating at the same bit rate per channel. 
Furthermore, since collisions with pulses from distant 
channels give the main contribution to energy shifts, 
a complete description of the dynamics must 
include interaction between pulses from all frequency channels. 
In contrast, effects of other nonlinear phenomena on pulse 
collisions are inversely proportional to some integer power 
of the frequency difference, and their cumulative influence 
can be adequately described by taking into account only a 
few neighboring channels \cite{Agrawal2001,Mollenauer99}.

Early studies of Raman cross talk in multichannel 
transmission systems focused on the dependence of the induced 
energy shifts on the total number of channels 
\cite{Chraplyvy84,Jander96}. The combined effects of  
Raman cross talk and randomness of pulse sequences were also 
considered, and it was found that the probability distribution 
function (PDF) of pulse amplitudes is lognormal 
\cite{Tkach95,Tkach97,Ho2000,Kumar2003}. However, these 
previous studies ignored several important properties of the system, 
which are essential for obtaining a correct dynamical model.  
First, all other nonlinear processes affecting pulse propagation, 
such as the Raman-induced self and cross frequency shifts were 
neglected. Second, strong coupling between amplitude dynamics and the 
dynamics of the other soliton parameters, such as frequency, 
position and phase, was not taken into account. Consequently, 
only the amplitude PDF was calculated, whereas a correct evaluation 
of system performance requires calculation of the mutual PDF of 
the pulse amplitude and position. Third, most studies considered 
only dynamic impact on performance of high frequency channels 
due to soliton decay, thus ignoring potential negative consequences 
for intermediate and low frequency channels due to large 
position shifts induced by relatively large amplitude values.

A more complete description of pulse propagation, 
which takes into account the three aforementioned 
factors, was developed in Refs. \cite{P2004,CP2005,P2007}. 
In Ref. \cite{P2004} it was shown that the coupling between 
frequency dynamics and amplitude dynamics leads to an  
exponential growth of the first two normalized moments of 
the self and cross frequency shifts with propagation distance. 
A perturbed NLS equation describing the 
combined effects of bit pattern randomness and Raman cross 
talk in a {\it two-channel} system was derived in 
Ref. \cite{CP2005}. Numerical simulations with the latter 
NLS model confirmed the analytic predictions of Ref. 
\cite{P2004}. Later on it was shown that the $n$th normalized moments 
of the self and cross frequency shifts increase exponentially 
with both propagation distance and $n^{2}$ \cite{P2007}. 
These results, combined with similar results for the normalized moments 
of the amplitude \cite{P2004}, imply that the soliton parameters 
exhibit intermittent dynamics in the sense that rare but 
violent events associated with relatively large amplitudes 
and frequency shifts become important. Furthermore, it was 
shown that the dominant mechanism for error generation 
in the system at long propagation distances is related to the intermittent 
dynamic behavior and is due to the Raman-induced cross frequency 
shift \cite{P2007}. In this process the error is generated due to 
large values of the frequency and position shifts induced by large 
amplitude values. Thus, it is very different from the two mechanisms 
for error generation that are usually considered in fiber optics 
transmission, which are due to: (1) position shift with almost 
constant amplitude, (2) amplitude decay with almost 
constant position shift. As mentioned above, the analysis  
in Ref. \cite{P2007} ignored radiation emission effects, 
which can be important in massive multichannel transmission. 
In the current paper we take these effects into account and 
derive a perturbed stochastic NLS model for propagation in a system 
with $N\ge 2$ frequency channels. We analyze the dynamics of 
the soliton parameters and the behavior of the BER by extensive 
numerical simulations with the model, and compare our results 
with the analytic calculations of Ref. \cite{P2007}.   

The material in the rest of the paper is organized as follows. 
In Sec. \ref{MF} A, we construct a perturbed stochastic NLS model 
describing soliton propagation in massive multichannel optical fiber 
transmission systems. The dynamics of the soliton amplitude and frequency 
and of the BER are obtained in Sec. \ref{MF} B by employing a standard 
adiabatic perturbation procedure. In Sec. \ref{simulation}, 
we analyze the results of numerical simulations with the perturbed
NLS model and compare them with the predictions of the adiabatic 
perturbation theory. Section \ref{conclusions} is reserved for 
conclusions.

\section{A stochastic model for pulse propagation}
\label{MF}
\subsection{Derivation of the model}
Propagation of short pulses of light through an optical fiber 
in the presence of delayed Raman response is described by 
the following perturbed NLS equation \cite{Agrawal2001}:
\begin{eqnarray}
i\partial_z\psi+\partial_t^2\psi+2|\psi|^2\psi=
-\epsilon_{R}\psi\partial_t|\psi|^{2},
\label{cfs1}
\end{eqnarray}
where $\psi$ is proportional to the envelope of the 
electric field, $z$ is propagation distance and $t$ 
is time in the retarded reference frame. The term 
$-\epsilon_{R}\psi\partial_t|\psi|^{2}$ accounts for 
the effects of delayed Raman response and $\epsilon_{R}$ 
is the Raman coefficient \cite{dimensions}.
When $\epsilon_{R}=0$, the single-soliton solution of 
Eq. (\ref{cfs1}) in a frequency channel $\beta$ 
is given by
\begin{eqnarray}
\psi_{\beta}(t,z)\!=\!
\eta_{\beta}\frac{\exp(i\chi_{\beta})}{\cosh(x_{\beta})},
\label{cfs2}
\end{eqnarray}
where $x_{\beta}=\eta_{\beta}\left(t-y_{\beta}-2\beta z\right)$, 
$\chi_{\beta}=\alpha_{\beta}+\beta(t-y_{\beta})+
\left(\eta_{\beta}^2-\beta^{2}\right)z$, 
and $\eta_{\beta}, \alpha_{\beta}$ and $y_{\beta}$ are the 
soliton amplitude, phase and position, respectively.

Consider the effects of delayed Raman response on a single collision
between two solitons from different frequency channels.
For simplicity, one of the two channels is chosen as the reference
channel with $\beta=0$ so that the frequency difference between the
two channels is $\beta$. We assume that $\epsilon_{R}\ll 1$ and
$1/|\beta|\ll 1$, which is the typical situation in
current multichannel transmission systems \cite{MM98}. 
In addition,  we assume that the two solitons are initially
well-separated from each other in the temporal domain.
Under these assumptions we can employ the perturbation procedure, 
developed in Refs. \cite{PCG2003,PCG2004,SP2004}, and applied in 
Ref. \cite{CP2005} for the case of delayed Raman response.
Here we only give the outline of the calculation and refer 
the interested reader to Ref. \cite{CP2005} for details.   
In accordance with this perturbative approach, we look 
for a two-pulse solution of Eq. (\ref{cfs1}) in the form
\begin{equation}
\psi_{two}=\psi_{0}+\psi_{\beta}+\phi,
\label{cfs3}
\end{equation}
where $\psi_{0}$ and $\psi_{\beta}$ are single-pulse 
solutions of Eq. (\ref{cfs1}) with $0<\epsilon_{R}\ll 1$
in channels 0 and $\beta$, respectively. 
The term $\phi$ on the right hand side of
Eq. (\ref{cfs3}) is a small  correction to the single-soliton
solutions, which is solely due to collision effects. By analogy 
with the ideal collision case we take $\phi$ to be of the form 
\begin{equation}
\phi=\phi_{0}+\phi_{\beta}+\dots
\,,
\label{cfs4}
\end{equation}
where $\phi_{0}$ and $\phi_{\beta}$ represent collision induced
corrections in channels $0$ and $\beta$, and the ellipsis
represents higher order terms in other channels. Combining 
Eqs. (\ref{cfs3}) and (\ref{cfs4}) we see that the total 
pulse in the reference channel is $\psi_{0}^{total}=\psi_{0}+
\phi_{0}$. We substitute the relations (\ref{cfs3}) and 
(\ref{cfs4}) together with 
$\psi_{0}(t,z)=\Psi_{0}(x_{0})\exp(i\chi_{0})$, 
$\phi_{0}(t,z)=\Phi_{0}(x_{0})\exp(i\chi_{0})$,
$\psi_{\beta}(t,z)=\Psi_{\beta}(x_{\beta})\exp(i\chi_{\beta})$,
and $\phi_{\beta}(t,z)=\Phi_{\beta}(x_{\beta})\exp(i\chi_{\beta})$ 
into Eq. (\ref{cfs1}). The resulting equation can be readily 
decomposed into an equation for the evolution of $\Phi_{0}$ 
and an equation for the evolution of $\Phi_{\beta}$. 
We focus attention on $\Phi_{0}$ and remark that the 
calculation of $\Phi_{\beta}$ is very similar. The equation 
for $\Phi_{0}$ is solved by integration with respect to 
$z$ over the collision region. Carrying out this integration 
one obtains that the $O(\epsilon_{R})$ effect of the collision 
on the reference channel soliton is given by 
\begin{eqnarray} &&
\Delta\Phi_{01}^{(1)}\!=
\eta_{\beta}\mbox{sgn}(\beta)\epsilon_{R}\Psi_{0}(x_{0}),
\label{cfs5}
\end{eqnarray}       
where the first subscript in $\Delta\Phi_{01}^{(1)}$ stands for the 
channel, the second subscript indicates the
combined order with respect to both $\epsilon_{R}$ and $1/\beta$, 
and the superscript represents the order in $\epsilon_{R}$. 
This $O(\epsilon_{R})$ effect corresponds to an amplitude 
change \cite{Chi89,Malomed91,Kumar98,CP2005} 
\begin{eqnarray}
\Delta\eta_{0}=
2\eta_{0}\eta_{\beta}\mbox{sgn}(\beta)\epsilon_{R},
\label{cfs6}
\end{eqnarray} 
which is accompanied by emission of continuous radiation. 
In a similar manner, one finds that the effect of the collision 
in order $\epsilon_{R}/\beta$ is \cite{CP2005}   
\begin{eqnarray} &&
\Delta\Phi_{02}^{(1)}\!=
\frac{4i\eta_{\beta}\epsilon_{R}}{|\beta|}\partial_t\Psi_{0}(x_{0}).
\label{cfs7}
\end{eqnarray}  
$\Delta\Phi_{02}^{(1)}$ corresponds to a collision induced 
frequency shift:
\begin{eqnarray}
\Delta\beta_{0}=-(8\eta_{0}^{2}\eta_{\beta}\epsilon_{R})/(3|\beta|),
\label{cfs8}
\end{eqnarray}
which is also accompanied by emission of continuous radiation. 

Let us describe propagation of a reference channel soliton 
under many collisions with solitons from all other frequency 
channels in a system with $2N+1$ channels. We employ a mean-field 
approximation \cite{CP2005}, in which we assume that the 
amplitudes of the solitons in the other channels are constant. 
The random character of soliton sequences in
different channels is taken into account by defining discrete
random variables $\zeta_{ij}$, which describe the occupation 
state of the $j$th time slot in the $i$th channel:
$\zeta_{ij}=1$ with probability $s$ if the slot is occupied, 
and 0 with probability $1-s$ otherwise. It follows that the 
$n$th moment of $\zeta_{ij}$ satisfies: 
$\langle\zeta_{ij}^{n}\rangle=s$. We also assume that the 
occupation states of different time slots are uncorrelated: 
$\langle\zeta_{ij}\zeta_{i'j'}\rangle=s^{2}$ if $i\ne i'$ and 
$j\ne j'$. We denote by $\Delta\beta$ the frequency   
difference between neighboring channels and by $T$ the time 
slot width. Therefore, the distance traveled by the reference 
channel soliton while passing two successive time slots in the 
nearby channels is $\Delta z_{c}^{(1)}=T/(2\Delta\beta)$. 
The $O(\epsilon_{R})$ effect of the collisions is taken into 
account by introducing a new perturbation term $S_{1}$ into 
Eq. (\ref{cfs1}). The term $S_{1}$ is obtained by summing 
Eq. (\ref{cfs5}) over all collisions occurring in the interval 
$\Delta z_{c}^{(1)}$, and dividing the 
result by $\Delta z_{c}^{(1)}$ 
\begin{eqnarray} &&
S_{1}\equiv 
i\epsilon_{R}\Psi_{0}e^{i\chi_{0}}
\sum_{i\ne 0}\mbox{sgn}(\beta_{i})
\!\!\sum_{j=(k-1)i+1}^{ki}\!\!
\frac{\zeta_{ij}}{\Delta z_{c}^{(1)}}, 
\label{cfs11}
\end{eqnarray}
where $k-1$ and $k$ are the indexes of the two successive time 
slots in the $i=-1$ channel, and the outside 
sum is from $-N$ to $N$. We decompose the disorder 
$\zeta_{ij}$ into an average part and a 
fluctuating part: $\zeta_{ij}=s+\tilde\zeta_{ij}$,
where $\langle\tilde\zeta_{ij}\rangle=0$, 
$\langle\tilde\zeta_{ij}\tilde\zeta_{i'j'}\rangle=s(1-s)
\delta_{ii'}\delta_{jj'}$, and $\delta_{ii'}$ is the 
Kronecker delta function. Substituting $\zeta_{ij}=s+\tilde\zeta_{ij}$ 
into Eq. (\ref{cfs11}) we obtain    
\begin{eqnarray} &&
\!\!\!\!\!\!\!S_{1}=
\frac{2is\epsilon_{R}\Delta\beta\psi_{0}}{T}
\sum_{i\ne 0}\mbox{sgn}(\beta_{i})|i|+
i\epsilon_{R}\xi(z)\psi_{0}, 
\label{cfs12}
\end{eqnarray}     
where the continuous disorder field $\xi(z)$ is
\begin{eqnarray} &&
\xi(z)=\frac{1}{\Delta z_{c}^{(1)}}
\sum_{i\ne 0}\mbox{sgn}(\beta_{i})
\sum_{j=(k-1)i+1}^{ki}\!\!\!\!\!\tilde\zeta_{ij}.
\label{cfs13}
\end{eqnarray} 
Using Eq. (\ref{cfs13}) and the properties of $\zeta_{ij}$ one 
can show that $\langle\xi(z)\rangle=0$ and 
$\langle\xi(z)\xi(z')\rangle=D_{N}\delta(z-z')$, 
where $D_{N}=N(N+1)D_{2}$, $D_{2}=2\Delta\beta s(1-s)T^{-1}$,   
and $\delta(z)$ is the Dirac delta function. Notice that the first 
term on the right hand side of Eq. (\ref{cfs12}) is zero due to
symmetry. Even if this term is not zero, it can be compensated 
by appropriately adjusting the gain of the amplifiers. Therefore, 
the $O(\epsilon_{R})$ effect of the collisions is described by
\begin{eqnarray} &&
S_{1}=i\epsilon_{R}\xi(z)\psi_{0}. 
\label{cfs14}
\end{eqnarray}        

The $O(\epsilon_{R}/\beta)$ effect of the collisions is calculated 
in a similar manner. We first sum Eq. (\ref{cfs7}) 
over all collisions occurring within the interval 
$\Delta z_{c}^{(1)}$:
\begin{eqnarray}&&
\!\!\!\!\!\!\!\!\!\!\!\!\!\!\tilde S_{2}\equiv
-c_{1}\partial_{t}\Psi_{0}
-4\epsilon_{R}\partial_{t}\Psi_{0}\sum_{i\ne 0}\frac{1}{|\beta_{i}|}
\!\!\sum_{j=(k-1)i+1}^{ki}\!\!
\frac{\tilde\zeta_{ij}}{\Delta z_{c}^{(1)}}, 
\label{cfs15}
\end{eqnarray}  
where $c_{1}=(16N\epsilon_{R}s)/T$. The second term on the 
right hand side of Eq. (\ref{cfs15}) can be estimated as 
$-8[D_{2}H_{N}/(T\Delta\beta)]^{1/2}\epsilon_{R}\partial_{t}\Psi_{0}$,   
where $H_{N}=\sum_{j=1}^{N}1/j$. Consequently, for a typical multichannel 
system the coefficient in front of $\epsilon_{R}\partial_{t}\Psi_{0}$ 
in this term is of order 1 or smaller, whereas for the first term 
this coefficient is of order $N$. We therefore neglect the 
second term on the right hand side of Eq. (\ref{cfs15}),  
and set $\tilde S_{2}=-c_{1}\partial_{t}\Psi_{0}$. 
Using the fact that for a weakly perturbed soliton 
$e^{i\chi_{0}}\partial_{t}\Psi_{0}=\partial_{t}\psi_{0}-
i\beta_{0}\psi_{0}$ we arrive at
\begin{eqnarray} &&
S_{2}\equiv
e^{i\chi_{0}}\tilde S_{2}=
-c_{1}\partial_{t}\psi_{0}+ic_{1}\beta_{0}\psi_{0}, 
\label{cfs16}
\end{eqnarray}
where $\beta_{0}$ is the frequency of the perturbed reference channel
soliton. Substituting $S_{1}$ and $S_{2}$ into 
Eq. ({\ref{cfs1}) and replacing $\psi_{0}$ with $\psi$ 
we obtain 
\begin{eqnarray} &&
i\partial_z\psi+\partial_t^2\psi+2|\psi|^2\psi=
-\epsilon_{R}\psi\partial_t|\psi|^{2}
\nonumber \\ &&
+i\epsilon_{R}\xi(z)\psi
-c_{1}\partial_{t}\psi+ic_{1}\beta_{0}(z)\psi,
\label{cfs17}
\end{eqnarray}
which is the stochastic model describing propagation of the 
reference channel soliton in the fiber under many collisions.

\subsection{Statistics of soliton parameters and BER calculation}
The evolution of the parameters of the reference channel soliton 
with propagation distance can be obtained by employing the standard 
adiabatic perturbation theory \cite{Kaup90,Hasegawa95}. 
Employing this perturbation procedure 
we obtain the following equations for the soliton amplitude 
and frequency:
\begin{eqnarray}&&
\frac{d\eta_{0}}{dz}=2\epsilon_{R}\xi(z)\eta_{0}(z),
\label{cfs21}
\end{eqnarray}
and 
\begin{eqnarray}&&
\frac{d\beta_{0}}{dz}=-\frac{8}{15}\epsilon_{R}\eta_{0}^{4}(z)
-\frac{2}{3}c_{1}\eta_{0}^{2}(z).
\label{cfs22}
\end{eqnarray}
Notice that the right hand side of Eq. (\ref{cfs21}) is contributed
solely by the second term on the right hand side of Eq. (\ref{cfs17}), 
i.e., the term describing the Raman cross talk effects. The first and 
second terms on the right hand side of Eq. (\ref{cfs22}) describe 
the Raman induced self- and cross-frequency shifts, and  
are contributed by the first and third terms on the right hand side of
Eq. (\ref{cfs17}), respectively. 

Integrating Eq. (\ref{cfs21}) over $z$ we obtain
\begin{eqnarray} &&
\eta_{0}(z)=\eta_{0}(0)\exp\left[2\epsilon_{R}x(z)\right],
\label{cfs23}
\end{eqnarray}  
where $x(z)=\int_{0}^{z}\mbox{d}z'\,\xi(z')$ and 
$\eta_{0}(0)$ is the initial amplitude. According to the central 
limit theorem, the PDF of $x(z)$ approaches a Gaussian PDF 
with $\langle x(z)\rangle=0$ and $\langle x^{2}(z)\rangle=D_{N}z$. 
As a result, the PDF of the soliton amplitude approaches 
a lognormal PDF: 
\begin{eqnarray}&&
\!\!\!\!\!\!\!\!\!\!\!\!\!\!\!\!\!F(\eta_{0})\!=\!
(8\pi D_{N}\epsilon_{R}^{2}z)^{-1/2}\eta_{0}^{-1}
\exp\!\left\{-\frac{\ln^{2}\left[\eta_{0}/\eta_{0}(0)\right]}
{8D_{N}\epsilon_{R}^{2}z}\right\}.
\label{cfs24}
\end{eqnarray}
The lognormal distribution is very different from the Gaussian 
distribution, and this difference is significant already in the 
main body of the distribution \cite{CP2005}. Moreover, the normalized 
moments of the lognormal PDF grow exponentially with propagation 
distance, from which it follows that the soliton amplitude 
exhibits intermittent dynamic behavior \cite{P2007}.  

The dynamic evolution of the soliton frequency is given by
\begin{eqnarray} &&
\beta_{0}(z)=\beta_{0}^{(s)}(z)+\beta_{0}^{(c)}(z),
\label{cfs25}
\end{eqnarray}
where 
\begin{eqnarray} &&
\beta_{0}^{(s)}(z)=
-\frac{8}{15}\epsilon_{R}\int_{0}^{z}\mbox{d}z'
\eta_{0}^{4}(z'), 
\label{cfs26}
\end{eqnarray}
is the self frequency shift and  
\begin{eqnarray} &&
\beta_{0}^{(c)}(z)=
-\frac{32N\epsilon_{R}s}{3T}\int_{0}^{z}\mbox{d}z'
\eta_{0}^{2}(z') 
\label{cfs27}
\end{eqnarray}
is the cross frequency shift. The $n$th moments of 
$\beta_{0}^{(s)}$ and $\beta_{0}^{(c)}$ can be calculated 
from \cite{P2007} 
\begin{eqnarray} &&
\!\!\!\!\!\!\langle\beta_{0}^{(s)n}(z)\rangle=
\left[-\frac{8}{15}\epsilon_{R}\eta_{0}^{4}(0)\right]^{n}n!
 \nonumber \\ &&
\!\!\!\!\!\times\prod_{m=1}^{n}
\int_{0}^{z_{m-1}}\!\!\!\! \mbox{d}z_{m}
\exp\left[32D_{N}\epsilon_{R}^{2}(2m-1)z_{m}\right], 
\label{cfs28}
\end{eqnarray} 
and 
\begin{eqnarray} &&
\!\!\!\!\!\!\langle\beta_{0}^{(c)n}(z)\rangle=
\left[-\frac{2}{3}c_{1}\eta_{0}^{2}(0)\right]^{n}n!
 \nonumber \\ &&
\!\!\!\!\!\times\prod_{m=1}^{n}
\int_{0}^{z_{m-1}}\!\!\!\! \mbox{d}z_{m}
\exp\left[8D_{N}\epsilon_{R}^{2}(2m-1)z_{m}\right], 
\label{cfs29}
\end{eqnarray} 
where $z_{0}=z$. By carrying out the integration in 
Eqs. (\ref{cfs28}) and  (\ref{cfs29})    
one can show that $\langle\beta_{0}^{(s)n}(z)\rangle$ 
and $\langle\beta_{0}^{(c)n}(z)\rangle$ are given by 
sums over exponential terms of the form 
$K_{m}\exp\left[a^{(s,c)}m^{2}D_{N}\epsilon_{R}^{2}z\right]$, 
where $a^{(s)}=32$, $a^{(c)}=8$, $0\le m\le n$, 
and the $K_{m}$ are constants. Furthermore, 
the leading contributions to the normalized moments 
$\langle\beta_{0}^{(s)n}(z)\rangle/\langle\beta_{0}^{(s)}(z)\rangle^{n}$ 
and $\langle\beta_{0}^{(c)n}(z)\rangle/\langle\beta_{0}^{(c)}(z)\rangle^{n}$ 
are exponentially increasing with both $z$ and $n^{2}$.     
As we shall see in the next Section, the normalized fourth 
moments of $\beta_{0}^{(s)}$, $\beta_{0}^{(c)}$ and $\beta_{0}$  
increase much faster with increasing propagation distance 
compared with the normalized second and third moments, which is  
a consequence of the intermittent nature of the dynamics.

In order to evaluate the system's BER we need to consider the 
main dynamical mechanisms leading to error generation. One 
mechanism, which has been widely studied in relation with Raman 
cross talk \cite{Tkach95,Chraplyvy84,Jander96,Tkach97,Ho2000,Kumar2003}, 
is due to pulse decay induced by loss of energy in 
collisions. This mechanism is associated with the 
small-$\eta$ tail of the amplitude PDF.      
Another mechanism, which has only recently been studied in 
relation with Raman cross talk, is due to the interplay between 
frequency (and position) dynamics and amplitude dynamics \cite{P2007}. 
In this case, the error is generated due to large values of the 
position shift, which are associated with large frequency shifts, 
and induced by relatively large values of the soliton amplitude.        
Notice that the lognormal statistics of the soliton amplitude 
leads to further enhancement of the BER contribution from the latter 
mechanism, since the large-$\eta$ tail of the lognormal PDF lies 
above the corresponding tail of the Gaussian PDF.
We are therefore interested in the soliton position shift, 
which is given by
\begin{eqnarray} &&
y_{0}(z)=y_{0}^{(s)}(z)+y_{0}^{(c)}(z),
\label{cfs31}
\end{eqnarray}
where
\begin{eqnarray} &&
y_{0}^{(s)}(z)=
-\frac{16\epsilon_{R}}{15}
\int_{0}^{z}\mbox{d}z'\int_{0}^{z'}\mbox{d}z''
\eta_{0}^{4}(z''), 
\label{cfs32}
\end{eqnarray}
and
\begin{eqnarray} &&
y_{0}^{(c)}(z)=
-\frac{64N\epsilon_{R}s}{3T}
\int_{0}^{z}\mbox{d}z'\int_{0}^{z'}\mbox{d}z''
\eta_{0}^{2}(z') 
\label{cfs33}
\end{eqnarray}
are the contributions from the self and cross frequency shifts, 
respectively. The position shift with a fixed amplitude 
$\eta_{0}(z)=\eta_{0}(0)=1$ is 
$\tilde y_{0}(z)=\tilde y_{0}^{(s)}(z)+\tilde y_{0}^{(c)}(z)$, 
where $\tilde y_{0}^{(s)}(z)= -(8\epsilon_{R}z^{2})/15$  
and $\tilde y_{0}^{(c)}(z)= -(32N\epsilon_{R}s z^{2})/(3T)$. 
The relative position shift is 
$\Delta  y_{0}(z)=\Delta  y_{0}^{(s)}(z)+\Delta  y_{0}^{(c)}(z)$, 
where $\Delta  y_{0}^{(s)}(z)=y_{0}^{(s)}(z)-\tilde y_{0}^{(s)}(z)$ 
and $\Delta  y_{0}^{(c)}(z)=y_{0}^{(c)}(z)-\tilde y_{0}^{(c)}(z)$.
We assume that $\tilde y_{0}$ can be compensated by 
employing filters. Therefore, the energy measured  
by the detector at a distance $z$ is 
\begin{eqnarray} &&
\!\!\!\!\!\!\!\!\!\!\!\!\!\!\!\!\!\!
I(\eta_{0},\Delta y_{0})= 
\eta_{0}^{2}\int_{-T/2}^{T/2}\!\!\mbox{d}t
\cosh^{-2}[\eta_{0}(t-\Delta y_{0})].
\label{cfs34}
\end{eqnarray}  
An occupied time slot is considered to be in error, if 
$I(\eta_{0},\Delta y_{0})\le I(z=0)/2\simeq 1$. 
We estimate the BER by numerically integrating Eqs. (\ref{cfs32}) 
and (\ref{cfs33}) coupled to Eq. (\ref{cfs23}) 
for different realizations of the disorder $\xi(z)$ 
and calculating the fraction of errored occupied time slots. 
The $z$-dependence of the BER obtained by this calculation 
is described in Sec. \ref{simulation}.

\section{Numerical simulations}
\label{simulation}
In the previous Section we calculated the statistics of the 
soliton parameters and the BER by employing the adiabatic 
perturbation theory and neglecting effects associated with 
emission of continuous radiation. We note that the latter 
effects can be particularly important for the system 
described by Eq. (\ref{cfs17}). Indeed, the second term on 
the right hand side of this equation has the form of 
disorder in the linear gain/loss coefficient. Such term 
can lead to instability with respect to emission 
of continuous radiation, which is of second order 
in $\epsilon_{R}$. It is therefore important to compare 
the results obtained in the previous Section  
by the reduced adiabatic method with results of numerical 
simulations with the more complete model, described by 
Eq. (\ref{cfs17}).   

Notice that the fourth term on the right hand side of 
Eq. (\ref{cfs17}) includes $\beta_{0}$. Since both 
$\beta_{0}$ and $c_{1}$ are of order $\epsilon_{R}$ this 
term is of order $\epsilon_{R}^{2}$, whereas the other 
perturbation terms in the equation are of order 
$\epsilon_{R}$. Moreover, since 
$\beta_{0}$ is a $z$-dependent random variable it is 
computationally complicated to solve Eq. (\ref{cfs17}) 
in its exact form. To overcome this problem, we replace 
$\beta_{0}$ in Eq. (\ref{cfs17}) with its value for 
the case where the amplitude is fixed and equal to 1: 
$\tilde\beta_{0}(z)=-(32N\epsilon_{R}s z)/(3T)-
(8\epsilon_{R}z)/15$. Thus, the perturbed NLS which 
we solve numerically is 
\begin{eqnarray} &&
i\partial_z\psi+\partial_t^2\psi+2|\psi|^2\psi=
-\epsilon_{R}\psi\partial_t|\psi|^{2}
\nonumber \\ &&
+i\epsilon_{R}\xi(z)\psi
-c_{1}\partial_{t}\psi+ic_{1}\tilde\beta_{0}(z)\psi.
\label{cfs40}
\end{eqnarray}
The initial condition is taken in the form of an ideal soliton:
$\psi(t,z=0)=\cosh^{-1}(t)$, with $\eta_{0}(0)=1$, $\beta_{0}(0)=0$,
$y_{0}(0)=0$, and $\alpha_{0}(0)=0$.

We perform Monte Carlo simulations with Eq. (\ref{cfs40}) 
with about $5\times 10^{4}$ disorder realizations. 
The equation is integrated by employing the split-step method 
with periodic boundary conditions. Numerical errors resulting
from radiation emission and the use of periodic boundary conditions
are overcome by applying artificial damping at the vicinity of the 
boundaries of the computational domain. The size of the domain is taken
to be $-100\leq t \leq 100$ so that the absorbing layers do
not affect the dynamics of the soliton pulses. The $t$-step and
$z$-step are taken as $\Delta t =0.048$ and $\Delta z =0.001$, 
respectively.

We focus attention on a transmission system with 101 channels 
operating at 10Gbits/s per channel. It should be emphasized that 
state-of-the-art experiments with dispersion-managed solitons 
demonstrated multichannel transmission with 109 channels 
at 10 Gbits/s per channel over a distance of 
$2\times 10^{4}$ km \cite{Mollenauer2003}. 
Several other experiments achieved total bit-rate 
capacities exceeding 1Tbits/s for shorter propagation distances 
\cite{Chraplyvy96,Inada2002,Grosz2004}.     
We use the following set of parameters, which is similar to the 
one used in multichannel transmission experiments with 
conventional solitons \cite{MM98}. 
Assuming that $T=5$, $\Delta\beta=10$, $s=1/2$ and 
$\eta_{i}(0)=1$ for all channels, the pulse width 
is 20 ps, $\epsilon_{R}=3\times 10^{-4}$, the channel spacing is 
75 GHz, and $D_{2}=1$. Taking $\beta_{2}=-1\mbox{ps}^{2}/\mbox{km}$,  
the soliton-peak-power is $P_{0}=1.25$ mW. For these values 
the width of the lognormal PDF in Eq. (\ref{cfs24}), 
which represents the strength of disorder effects, is 
$8D_{N}\epsilon_{R}^{2}z=1.8\times 10^{-3}z$ 
for the reference channel. For $z=25$, corresponding to
transmission over $2\times 10^{4}$ km,     
$8D_{N}\epsilon_{R}^{2}z=0.046$.

The $z$-dependences of the $n=2,3,4$ normalized moments of 
$\beta_{0}^{(s)}$, $\beta_{0}^{(c)}$, and $\beta_{0}$ 
as obtained by numerical solution of the perturbed NLS 
are shown in Fig. \ref{fig1} together with the results of 
the adiabatic perturbation theory. The numerical simulation 
results for $\beta_{0}^{(s)}$ and  $\beta_{0}^{(c)}$ were obtained 
by solving the reduced models
\begin{eqnarray} &&
\!\!\!\!\!\!\!\!\!\!\!
i\partial_z\psi+\partial_t^2\psi+2|\psi|^2\psi=
-\epsilon_{R}\psi\partial_t|\psi|^{2}
+i\epsilon_{R}\xi(z)\psi,
\label{cfs41}
\end{eqnarray}
and
\begin{eqnarray} &&
i\partial_z\psi+\partial_t^2\psi+2|\psi|^2\psi=
i\epsilon_{R}\xi(z)\psi
\nonumber \\ &&
-c_{1}\partial_{t}\psi+ic_{1}\tilde\beta_{0}(z)\psi
\label{cfs42}
\end{eqnarray}
with $\tilde\beta_{0}(z)=-(32N\epsilon_{R}s z)/(3T)$, 
respectively. The results obtained by numerical solution 
of the perturbed NLS equation are in good agreement 
with those obtained by the adiabatic perturbation theory.       
Moreover, one can see that the fourth moments of 
$\beta_{0}^{(s)}$, $\beta_{0}^{(c)}$, and $\beta_{0}$  
increase much faster with increasing $z$ compared with 
the second and third moments, in 
accordance with the intermittent nature of the dynamics. 
In addition, the normalized moments of $\beta_{0}^{(s)}$ 
grow faster than those of $\beta_{0}^{(c)}$. This can be explained 
by noting that the rate of change of $\beta_{0}^{(s)}$ 
is proportional to $\eta_{0}^{4}$, whereas $d\beta_{0}^{(c)}/dz$ 
is proportional to $\eta_{0}^{2}$. Notice, however, that the 
values of the normalized moments of the total frequency shift 
$\beta_{0}$ are very close to those of $\beta_{0}^{(c)}$. 
This is due to the fact that for the system described above 
$\beta_{0}^{(c)}$ is typically much larger than $\beta_{0}^{(s)}$.     
\begin{figure}[ptb]
\begin{tabular}{cc}
\epsfxsize=6.2cm  \epsffile{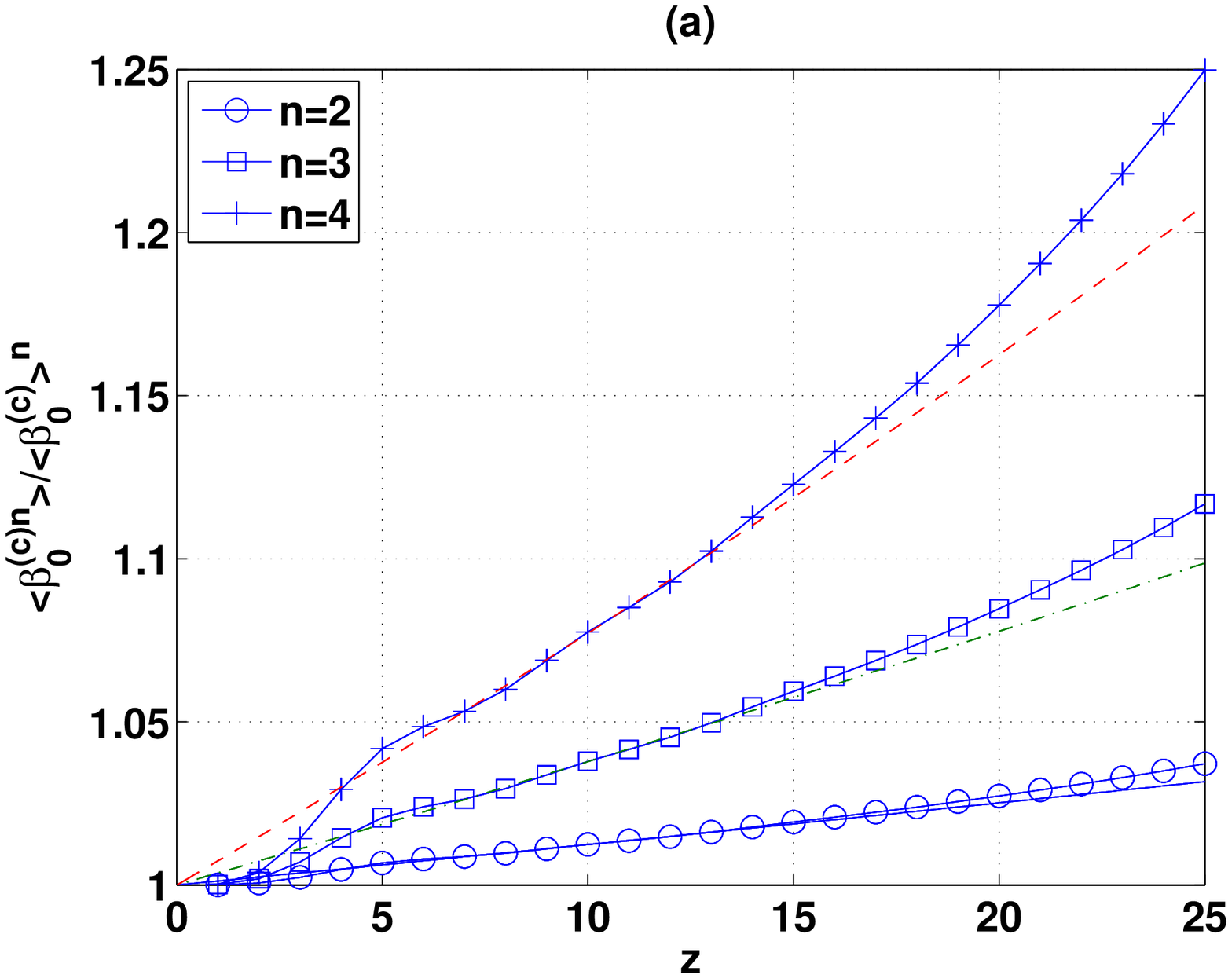}\\
\epsfxsize=6.2cm  \epsffile{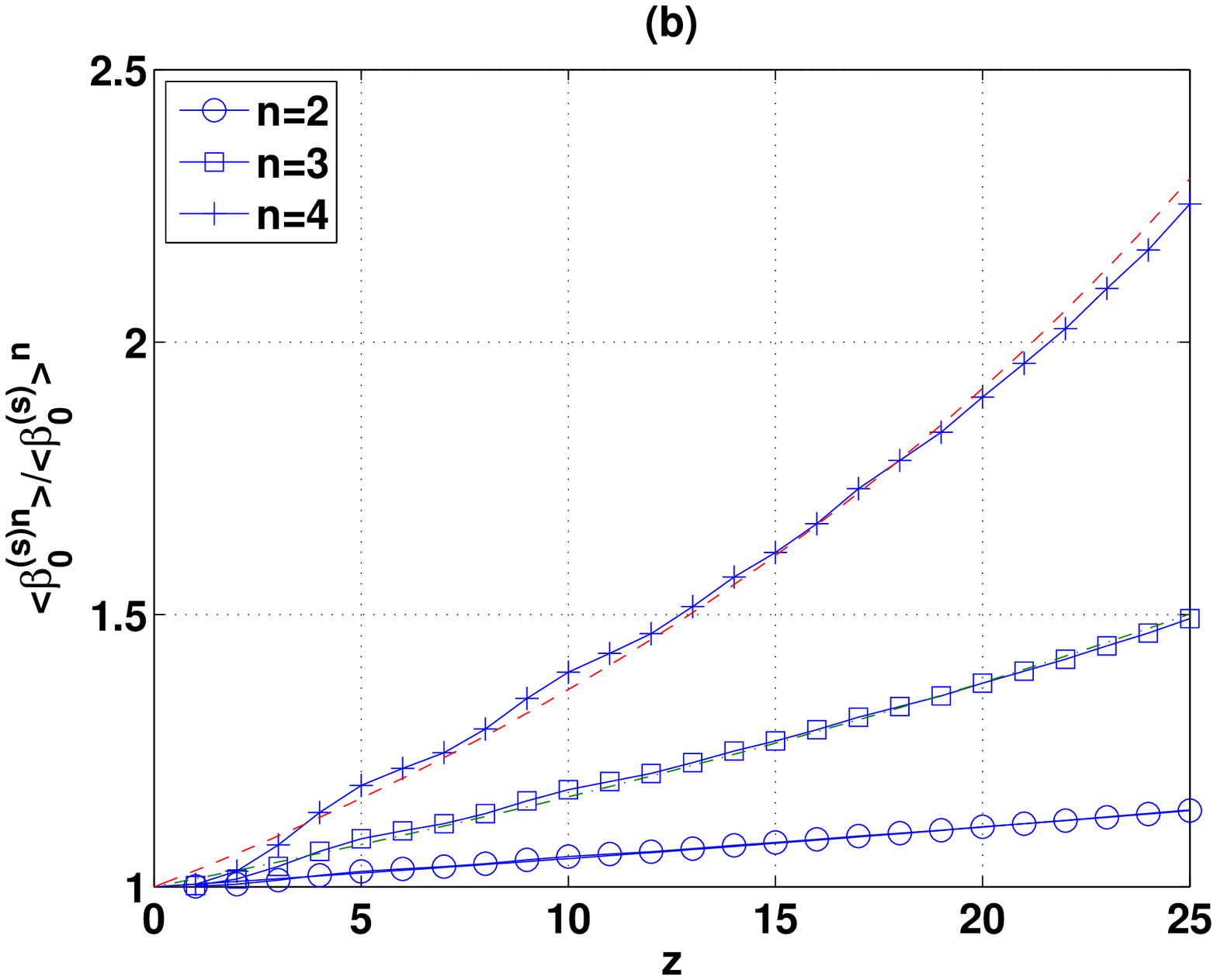}\\
\epsfxsize=6.2cm  \epsffile{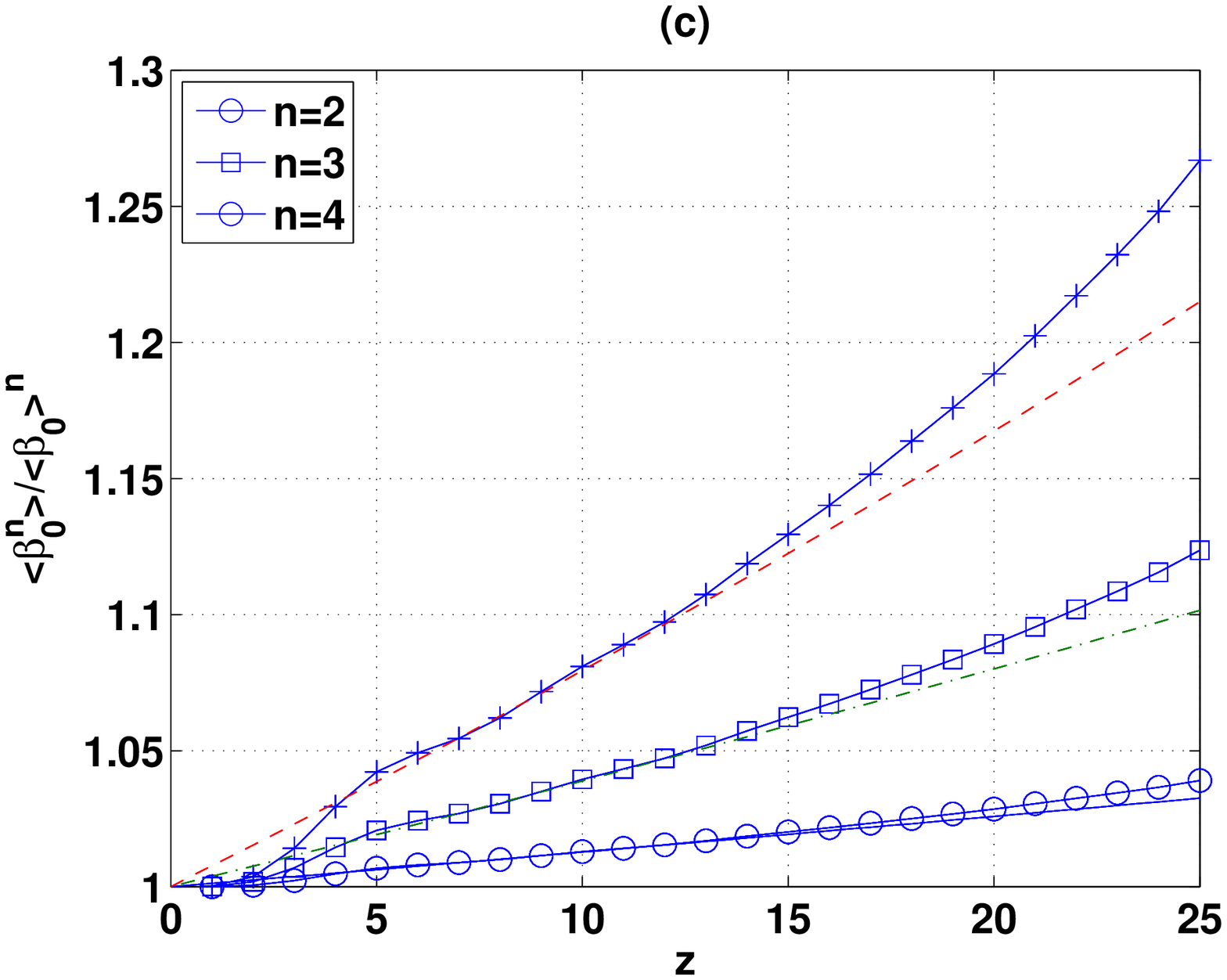}
\end{tabular}
\caption{Normalized moments of the reference channel soliton's   
cross frequency shift (a), self frequency shift (b), and total 
frequency shift (c) vs propagation distance $z$ for 
a multichannel system with 101 channels at 10 Gbits/s per channel.
The solid, dashed-dotted, and dashed lines correspond to the 
$n=2,3,4$ normalized moments obtained by the adiabatic 
perturbation method, using Eqs. (\ref{cfs27}) and (\ref{cfs29}) 
in (a), Eqs. (\ref{cfs26}) and (\ref{cfs28}) in (b), and Eqs. 
(\ref{cfs25})-(\ref{cfs27}) in (c). 
The circles, squares and crosses represent the 
$n=2,3,4$ normalized moments obtained by numerical integration 
of Eq. (\ref{cfs42}) in (a), Eq. (\ref{cfs41}) in (b), and 
Eq. (\ref{cfs40}) in (c).}
\label{fig1}
\end{figure}

The BER of the reference channel is calculated by the procedure 
described in Sec. II B. That is, we calculate the measured intensity 
using Eq. (\ref{cfs34}), and declare an occupied time slot to be 
in error, if $I(\eta_{0},\Delta y_{0})\le I(z=0)/2\simeq 1$. 
The $z$-dependence of the BER obtained by numerical integration 
of Eq. (\ref{cfs40}) is shown in Fig. \ref{fig2} together with the result 
obtained by employing the adiabatic perturbation procedure. 
The agreement between the perturbed NLS simulations and the 
corresponding adiabatic theory calculations  
is good. Furthermore, it is seen that the BER attains relatively 
large values, which range from about $3\times 10^{-5}$ for 
$z=16$ ($X=1.28\times 10^{4}$ km) to about $10^{-1}$ at 
$z=25.0$ ($X=2\times 10^{4}$ km).
We remark that the BER values obtained by integrating 
Eq. (\ref{cfs42}), which takes into account only $y_{0}^{(c)}$, 
are very close to the ones obtained by solving Eq. 
(\ref{cfs40}), which takes into account 
both $y_{0}^{(s)}$ and $y_{0}^{(c)}$. In fact, since the difference between 
the two BER curves is indistinguishable on the scale of Fig. \ref{fig2}, 
we choose to omit the result obtained with Eq. (\ref{cfs42}).    
The fact that the two models [Eq. (\ref{cfs40}) and Eq. (\ref{cfs42})] 
give such close BER values is explained by noting that the 
cross frequency shift is typically much larger than 
the self frequency shift for the multichannel system 
considered here.   
\begin{figure}[ptb]
\epsfxsize=7.5cm  \epsffile{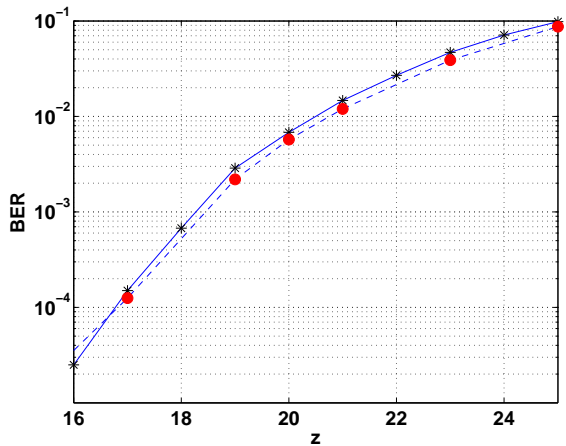}
\caption{The $z$ dependence of the BER for the reference channel 
in a 101-channel transmission system operating at 10Gbits/s per channel. 
The circles represent the adiabatic perturbation prediction 
obtained by using Eqs. (\ref{cfs23}), (\ref{cfs32}), and (\ref{cfs33}), 
while the stars correspond to the result of numerically 
integrating Eq. (\ref{cfs40}).}
\label{fig2}
\end{figure}

As explained in Section \ref{MF} and in Ref. \cite{P2007}, 
the two main mechanisms for error generation in the multichannel 
system are: (1) pulse decay, (2) large position shifts. While 
the first mechanism is associated with small pulse amplitudes, 
the second one is predominantly due to large amplitudes. 
Hence, in order to better understand the roles of these 
two error-generating mechanisms one has to study the mutual PDF 
$G(\eta_{0},\Delta y_{0})$. Figure \ref{fig3} shows 
$G(\eta_{0},\Delta y_{0})$ for the reference channel 
soliton at the final propagation distance $z=25$. The result   
obtained by numerical solution of Eq. (\ref{cfs40})
(Fig. \ref{fig3}(a)) is in good agreement with the prediction of 
the adiabatic perturbation theory (Fig. \ref{fig3}(b)). Moreover, 
the mutual distribution function is strongly asymmetric about the
$\Delta y_{0}=0$ and $\eta_{0}=1$ axes. This 
asymmetry is a direct consequence of the strong coupling between position 
dynamics and amplitude dynamics, as can be seen from 
Eqs. (\ref{cfs32}) and (\ref{cfs33}). We emphasize that 
this behavior is very different from the one observed for 
soliton propagation in single-channel systems in the presence 
of amplifier noise, where the mutual PDF is approximately symmetric 
with respect to both $\Delta y_{0}=0$ and $\eta_{0}=1$ 
\cite{Falkovich2001,Falkovich2004}. We also note that the mutual 
PDF shown in Fig. \ref{fig3} is skewed toward larger $\eta_{0}$ 
values, which can be explained by the skewed character of the 
lognormal distribution $F(\eta_{0})$.

\begin{figure}[ptb]
\epsfxsize=7.5cm  \epsffile{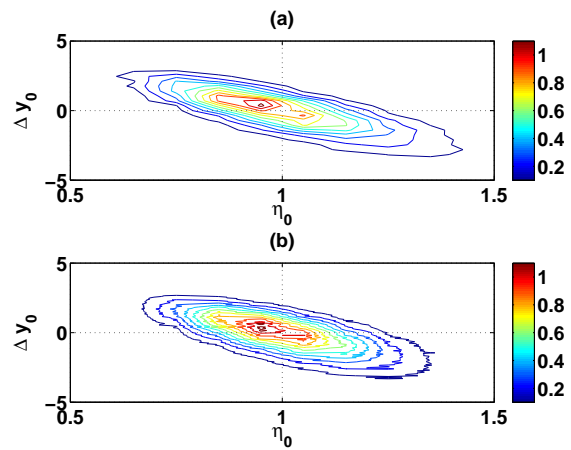}
\caption{The mutual PDF $G(\eta_{0},\Delta y_{0})$ 
for the reference channel soliton at $z=25$ as obtained by numerical 
integration of Eq. (\ref{cfs40}) (a), and as predicted by the adiabatic 
perturbation method (b).} 
\label{fig3}
\end{figure}

\section{Conclusions}
\label{conclusions}
We investigated propagation of optical pulses in massive multichannel 
optical fiber communication systems, taking into account the effects 
of delayed Raman response and bit pattern randomness. We derived a 
mean-field description of the propagation, which is given by a 
perturbed stochastic NLS equation that takes into account changes in pulse 
energy and momentum. This perturbed NLS model includes 
the effects of emission of continuous radiation, which were neglected 
in the simpler adiabatic perturbative approach used in Ref. 
\cite{P2007}. Our numerical simulations show that the normalized 
moments of the soliton frequency shift grow exponentially with 
propagation distance. Furthermore, the dynamics leads to relatively 
high values of the BER at intermediate and large propagation 
distances, and to an asymmetric form of the mutual 
PDF of amplitude and position. These results are 
in good agreement with results of the adiabatic perturbation theory. 
We therefore conclude that the interplay between Raman scattering and 
bit pattern randomness plays a very important role in massive 
multichannel transmission systems. 

The exponential growth of the normalized moments of soliton parameters 
is indicative of intermittent dynamics in the following sense: for certain 
realizations of pulse sequences (in the other frequency channels) the 
reference channel soliton can experience relatively large changes in its 
amplitude, which lead to relatively large position shifts and BER values. 
This dynamic behavior is quite surprising, since intermittency is 
usually associated with strongly nonlinear systems such as turbulence 
and chaotic flow \cite{Frisch95}, whereas optical fiber systems are 
only weakly nonlinear. In a recent paper \cite{P2007b}, one of the 
authors suggested that this unexpected similarity might not be coincidental,  
but rather a consequence of the similarity between the dynamic behavior of  
the soliton amplitude in the fiber optics system and the 
behavior of the local average of energy dissipation 
in turbulent flow.    

We conclude by noting that the dynamic behavior described in 
the current paper is not limited to conventional optical solitons. 
Indeed, it is known that the effects of delayed Raman response on a single 
collision between two dispersion-managed (DM) solitons are very similar 
to the effects in the conventional soliton case \cite{Kaup99}. 
Therefore, we expect that similar results would hold for  
DM multichannel transmission systems as well. A different type of 
nonlinearity that can lead to similar dynamics is due to nonlinear 
loss/gain. In this case pulse propagation is described by 
a perturbed NLS equation, in which the 
$-\epsilon_{R}\psi\partial_t|\psi|^{2}$ term is replaced 
by $\mp\epsilon_{c}|\psi|^{2}\psi$, where 
$\epsilon_{c}$ is the cubic nonlinear loss/gain coefficient.    
It can be shown that the main effect of a fast collision in the 
presence of nonlinear loss/gain is a change in the soliton 
amplitude, which is given by an equation of the form 
(\ref{cfs6}) with $\mbox{sgn}(\beta)\epsilon_{R}$ replaced 
by $\mp 2\epsilon_{c}/|\beta|$. If additional perturbations that 
affect the soliton frequency and position exist, the dynamics 
of the frequency or position will usually be coupled to the 
amplitude dynamics in a manner similar to the one described 
in Section II. Consequently, our results should also be 
applicable for propagation of NLS solitons in systems with 
nonlinear loss or gain.


\begin{thebibliography}{}
\bibitem{Malomed89} Y. S. Kivshar and B. A. Malomed,   
Rev. Mod. Phys. {\bf 61}, 763 (1989).  

\bibitem{Frisch95} U. Frisch, {\it Turbulence: The Legacy 
of A. N. Kolmogorov}, (Cambridge University Press, Cambridge, 1995).

\bibitem{Agrawal2001} G. P. Agrawal, {\it Nonlinear 
Fiber Optics} (Academic, San Diego, CA, 2001).

\bibitem{Menyuk95} C. R. Menyuk, Opt. Lett. {\bf 20}, 285 (1995).

\bibitem{Georges96} T. Georges, Opt. Commun. {\bf 123}, 617 (1996).   

\bibitem{Falkovich2001} G. E. Falkovich, I. Kolokolov, V. Lebedev, 
and S. K. Turitsyn, Phys. Rev. E {\bf 63}, 025601 (2001).

\bibitem{Biondini2002} G. Biondini, W. L. Kath, and C. R. Menyuk, 
IEEE Photon. Technol. Lett. {\bf 14}, 310 (2002).  

\bibitem{Falkovich2004} G. Falkovich, I. Kolokolov, V. Lebedev, 
V. Mezentsev, and S. Turitsyn, Physica D {\bf 195}, 1 (2004).

\bibitem{P2007} A. Peleg, Phys. Lett. A {\bf 360}, 533 (2007).

\bibitem{Mitschke86} F. M. Mitschke and L. F. Mollenauer,
Opt. Lett. {\bf 11}, 659 (1986).

\bibitem{Gordon86} J. P. Gordon, Opt. Lett. {\bf 11}, 662 (1986).

\bibitem{Kodama87} Y. Kodama and A. Hasegawa, 
IEEE J. Quantum Electron. {\bf QE-23}, 510 (1987). 

\bibitem{Chi89} S. Chi and S. Wen, Opt. Lett. {\bf 14}, 1216 (1989).

\bibitem{Malomed91} B. A. Malomed, Phys. Rev. A {\bf 44}, 1412 (1991).

\bibitem{Stolen92} R. H. Stolen and W. J. Tomlinson, 
J. Opt. Soc. Am. B {\bf 9}, 565 (1992).
 
\bibitem{Agrawal96} C. Headley III and G. P. Agrawal, 
J. Opt. Soc. Am. B {\bf 13}, 2170 (1996).

\bibitem{Kumar98} S. Kumar, Opt. Lett. {\bf 23}, 1450 (1998). 

\bibitem{Kaup99} T. I. Lakoba and D. J. Kaup, Opt. Lett. {\bf 24},
  808 (1999).

\bibitem{Chung2002} F. G. Omenetto, Y. Chung, D. Yarotski, 
T. Schaefer, I. Gabitov, and A. J. Taylor, 
Opt. Commun. {\bf 208}, 191 (2002).

\bibitem{Skryabin2003} D. V. Skryabin, F. Luan, J. C. Knight, 
and P. S. Russell, Science {\bf 301}, 1705 (2003).  
  
\bibitem{Skryabin2006} F. Luan, D. V. Skryabin, A. V. Yulin, 
and J. C. Knight, Opt. Express {\bf 14}, 9844 (2006).

\bibitem{Bang2006}  M. H. Frosz, O. Bang, and A. Bjarklev, 
Opt. Express {\bf 14}, 9391 (2006).    

\bibitem{Islam2004} M. N. Islam, ed.,  
{\it Raman Amplifiers for Telecommunications 1: Physical Principles}  
(Springer, New York, 2004).

\bibitem{Agrawal2005} C. Headley and G. P. Agrawal, eds., 
{\it Raman Amplification in Fiber Optical Communication Systems} 
(Elsevier, San Diego, CA, 2005).

\bibitem{Stolen72} R. H. Stolen, E. P. Ippen, and A. R. Tynes, 
Appl. Phys. Lett. {\bf 20}, 62 (1972). 

\bibitem{Tkach97} F. Forghieri, R. W. Tkach, and A. R. Chraplyvy,
 in {\it Optical Fiber Telecommunications III}, I. P. 
Kaminow and T. L. Koch, eds., (Academic, San Diego, CA, 1997), 
Chapter 8, Sec. VIII.

\bibitem{Mollenauer99} P. V. Mamyshev and L. F. Mollenauer, 
Opt. Lett. {\bf 24}, 448 (1999).

\bibitem{Chraplyvy84}  A. R. Chraplyvy, Electron. Lett. {\bf 20}, 
58 (1984).

\bibitem{Jander96} D. N. Christodoulides and R. B. Jander, 
IEEE Photon. Technol. Lett. {\bf 8}, 1722 (1996).    

\bibitem{Tkach95} F. Forghieri, R. W. Tkach, and A. R. Chraplyvy,
IEEE Photon. Technol. Lett. {\bf 7}, 101 (1995).

\bibitem{Ho2000} K. P. Ho, J. Lightwave Technol. {\bf 18}, 
915 (2000). 

\bibitem{Kumar2003} M. Muktoyuk and S. Kumar, 
IEEE Photon. Technol. Lett. {\bf 15}, 1222 (2003).     

\bibitem{P2004} A. Peleg, Opt. Lett. {\bf 29}, 1980 (2004).

\bibitem{CP2005} Y. Chung and A. Peleg, Nonlinearity {\bf 18}, 
1555 (2005). 

\bibitem{dimensions} The dimensionless $z$ in Eq. (\ref{cfs1}) is
$z=(|\beta_{2}|X)/(2\tau_{0})$, where $X$ is the actual position,
$\tau_{0}$ is the soliton width, and $\beta_{2}$ is the second order
dispersion coefficient. The dimensionless retarded time is
$t=\tau/\tau_{0}$, where $\tau$ is the retarded time.  
The spectral width is $\nu_{0}=1/(\pi^{2}\tau_{0})$ and the 
frequency difference is $\Delta\nu=(\pi\Delta\beta\nu_{0})/2$. 
$\psi=E/\sqrt{P_{0}}$, where $E$ is proportional to the 
electric field and $P_{0}$ is the peak power. 
The dimensionless second order dispersion coefficient is 
$d=-1=\beta_{2}/(\gamma P_{0}\tau_{0}^{2})$, 
where $\gamma$ is the Kerr nonlinearity coefficient.
The coefficient $\epsilon_{R}$ is given by 
$\epsilon_{R}=0.006/\tau_{0}$, where $\tau_{0}$ 
is in picoseconds.

\bibitem{MM98} L. F. Mollenauer and P. V. Mamyshev,
IEEE J. Quantum Electron. {\bf 34}, 2089 (1998).

\bibitem{PCG2003} A. Peleg, M. Chertkov, and I. Gabitov, 
Phys. Rev. E {\bf 68}, 026605 (2003).

\bibitem{PCG2004} A. Peleg, M. Chertkov, and I. Gabitov,  
J. Opt. Soc. Am. B {\bf 21}, 18 (2004).

\bibitem{SP2004} J. Soneson and A. Peleg, 
Physica D {\bf 195}, 123 (2004).    

\bibitem{Kaup90} D. J. Kaup, Phys. Rev. A {\bf 42}, 5689 (1990).

\bibitem{Hasegawa95} A. Hasegawa and Y. Kodama,   
{\it Solitons in Optical Communications} 
(Clarendon, Oxford, 1995). 
  
\bibitem{Mollenauer2003} L. F. Mollenauer, A. Grant, X. Liu, 
X. Wei, C. Xie, and I. Kang, Opt. Lett. {\bf 28}, 2043 (2003). 

\bibitem{Chraplyvy96}  A. R. Chraplyvy , A. H. Gnauck , R. W. Tkach, 
J. L. Zyskind, J. W. Sulhoff, A. J. Lucero, Y. Sun, R. M. Jopson, 
F. Forghieri, R. M. Derosier, C. Wolf, and A. R. McCormick,   
IEEE Photon. Technol. Lett. {\bf 8}, 1264 (1996).


\bibitem{Inada2002} Y. Inada, H. Sugahara, K. Fukuchi, T. Ogata, 
Y. Aoki, IEEE Photon. Technol. Lett. {\bf 14}, 1366 (2002).

\bibitem{Grosz2004} D. F. Grosz, A. Agarwal, S. Banerjee, D. N. Maywar, 
A. P. K\"ung, J. Lightwave Technol. {\bf 22}, 423 (2004). 

\bibitem{P2007b} A. Peleg, ``Raman cross talk between optical 
solitons as a random cascade model'', eprint arXiv:0706.4333.     

\end{thebibliography}
\end{document}